\documentclass[12pt]{article}

\usepackage{scicite}

\usepackage{times}
\usepackage{url}
\usepackage{graphicx}
\usepackage{graphicx}
\usepackage{subfigure}
\usepackage{caption}
\usepackage{mathrsfs}
\usepackage{amsmath}
\usepackage{amssymb}
\usepackage{array} 
\newenvironment{sciabstract}{%
\begin{quote} \bf}
{\end{quote}}

\newcounter{lastnote}

\title{Protocol Programming: A Connection \\ of the Digital World}

\author
{Yanping Chen,$^{1\ast}$ Qinghua Zheng,$^{1}$ Ping Chen$^{2}$\\
\\
\normalsize{$^{1}$Xi'an Jiaotong University, China}\\
\normalsize{$^{2}$University of Massachusetts Boston, USA}\\
\normalsize{E-mail: ypench@gmail.com, qhzheng@mail.xjtu.edu.cn, ping.chen@umb.edu}
}
\date{}

\begin{document} 


\baselineskip18pt

%\baselineskip24pt


\maketitle

\begin{sciabstract}
The current computer programmings encapsulate attributes and behaviours into objects, but miss the mechanism to support the connection among objects. A programming paradigm is presented to connect all objects. The connection supports communications. Protocols are defined to coordinate the behaviours between objects, which enable the interaction of objects across different platforms. The connection also provides an efficient mechanism to support the concurrency, parallelism, distribution, pipeline and adaptability, etc. They can be governed transparently, autonomously, even adaptively. In this paper, an implementation is also discussed to show the effectiveness of protocol programming.
\end{sciabstract}

\section*{Connection is the norm}
\label{sec:introduction}

The connection between entities is an important aspect of the world. Large objects as celestial bodies or small objects as biomolecules are connected, by gravity or conjugation, and organized into a galaxy or an organism. The connection should not be treated as an attribute of an object. It has a powerful influence on the connected objects. For example, gravity controls trajectories of celestial bodies, chemical reactions determine the transformation between substances. And relations, e.g., food chain, heredity, etc. judge the fittest and influence the evolution of species. The connection condenses substances into the cosmos, synthesizes amino acid and brings the life. All things are influenced or developed by connecting with others.

In this article, we define ``connection'' as the channel for interactions between objects. Usually interactions are not random, they show a clear pattern. The patterned interaction is referred to as ``communication'' in this paper. The connection supports communications between objects. It is the core idea of our method, we give an example for easier understanding. Let's consider the ocean, the land and the sky as a platform for the the biological world. Animals wander around the platform. It should be emphasized that the ability for an animal to ramble is important, referred to as ``autonomy''. It enables animals to change to fit its surroundings. Strolling animals are connected by inhabiting or encountering with each others. In the wild, passing each other does nothing about the communication. The communication between them is established by giving birth to, feeding, preying, fighting, etc.

In the digital world, to imitate or encode a system under a programming paradigm, the modularity is an important issue, which supports structured programming. Modules (e.g., classes or functions \cite{hudak1989conception,armstrong2006quarks}) are used to encapsulate objects and partition a system into different parts. However, in practice, it is common that a complex problem can not be perfectly partitioned. For example, to design a building, some parts of a building (e.g., the water supply or power system) are difficult to be encapsulated. This is also known as the crosscutting issue \cite{elrad2001discussing}. It makes the code tangled and increases the cost of communication among developers when designing and developing a system \cite{brooks1975mythical}. Furthermore, in object-oriented programming, an object is merely nested in another object as a member variable. The inner object is restricted by the holder that difficult to be shared with others. It also sacrifices the autonomy of an object, making it difficult to update independently.

Without a mechanism to support the connection appropriately, a hard partition hurts the communications between objects. In the traditional programming paradigms, various approaches are applied to support communications between objects. For example, objects can be used as function parameters or member variables of other objects. Inheritance is used to represent relations between classes. It does nothing about the communication unless static members are defined in the declaration \cite{wirfs1990surveying}. Then instantiated objects can share the same variables for communications. Others such as message passing, semaphore, shared memory, socket, etc. are widely adopted to support communication. Because the lack of the connection between objects, these approaches are used to support the communication weakened by separating a system into discrete parts. Traditionally, the connection as a mechanism to support communications has received less attention. The communications between objects are temporary and irregular, mainly used to exchange data or control resource competition if required. It is rarely formalized as a connection mechanism to support communications.

Compared to the traditional paradigms, connection in protocol programming is developed into a new territory. Under the protocol programming, programming is divided into two steps: protocol implementation and domain implementation. Protocol implementation builds an infrastructure to support the communication between objects. Protocols are defined to control the communication between them. By inheritance from a protocol implementation all instantiated objects are automatically connected into a network. Domain implementation is developed by traditional programming paradigms, encapsulating attributes and behaviours into objects. While supported by protocol implementation, partitioning a system into discrete models and reorganizing them become easier and more effective.

\section*{Connection creates possible}
\label{sec:motivation}

Building a system is usually not a trivial matter. It is often partitioned into discrete modules (e.g., procedures, functions, objects, agents, aspect, etc.), then a team is set up to implement each module separately. It is generally accepted that a modular design is the key to successful programming. All programming paradigms have mechanisms to support structured programming, where the modularity is guaranteed. Systems developed by structured programming have good stability, efficiency, maintainability and expansibility. The main obstacle for structured programming is that some problems are difficult to be partitioned. In some projects, design decision is scattered throughout a system. A hard partition breaks the connections between modules and hurts the communications between them.

In the digital world, to support the partition and connection between modules, many program paradigms are developed, e.g., agent-oriented programming \cite{shoham1993agent}, aspect-oriented programming \cite{kiczales1997aspect}. Based on message passing, mechanisms are proposed for communicating between threads or processes (e.g., Message Passing Interface \cite{gropp1996high}, Erlang \cite{armstrong1993concurrent}, Hoare et al. \cite{hoare1978communicating}). The communication across different systems are also designed (e.g., MapReduce \cite{dean2008mapreduce}, service-oriented computing \cite{papazoglou2003service}, CORBA \cite{siegel1998omg}). In a program, mechanisms (e.g., semaphore, shared memory, socket, message passing) are used to support the communication between partitioned modules. Because implementation of these mechanisms is not transparent to programmers, it may lead to many issues such as mutual exclusion, interrupts, data conflict, resource competition, etc. When building a system, these mechanisms increase the burden of design and implementation, and make the code difficult to understand and maintain. Furthermore, when a system is complex enough, partitioning it, defining the interfaces and integrating every part are very challenging. The cost to design, develop and maintain such a system and the communication between developers are increased considerably \cite{brooks1975mythical}.

To better illustrate our ideas, we analyse two systems (the Internet and the biosphere), which show the motivation of protocol programming.

Since the invention of packet switching theory \cite{kleinrock1961information}, for half a century, the Internet has expanded into a huge system with billions of access points and numerous applications \cite{leiner2009brief}. At the dawn of computer communication \cite{marill1966toward}, it is difficult to foresee that human society can build such a complex system. Despite various techniques are developed to accelerate the growth of the Internet, the most important technique is the Internet protocol deigned for the open-architecture network environment. Under the internet protocols, every equipment is independently connected into the Internet and the information between them are routed automatically. The Internet protocol hides the underneath implementation of connected modules. By following the Internet protocol, each part of the Internet can conveniently interact with each other across heterogeneous platforms or systems. Moreover, because the Internet protocol is a connectionless service, each part can be updated independently.

The biosphere is a largely self-regulated system. The earth can be seen as a platform for life to compete. Individual organisms of the earth encounter on the land, in the air or the water. The connection (encounter) is an important aspect for life on the earth. After the connection was constructed, the interaction between them is not random. They follow inherent patterns (e.g., foster, prey, competition, hereditary and variation) in the same way as the protocol. Life in the biosphere is ruled by these patterns, referred to as laws in biology. Hereditary and variation are two important laws determining the survival of a species and the evolution of the living nature. Many factors can disturb the natural stability of the biosphere, e.g., the geographical or climatic factor, the species diversity and the food chain, etc. They are the motive force for the species variation. The variation enables that a species can adapt to the environmental change. ``survival of the fittest'' ensures that the suitable change is inherited. Accumulated changes in individuals or species give rise to the evolution of the species.

Designing or constructing a system like the Internet or the biosphere in one step is far beyond human ability. They are not built, but evolved. Five characteristics can be induced from such systems. Each system should consist of modules or individuals that are mutually independent, which enables the {\em separableness}. Individuals can interact with each other and the {\em connection} is guaranteed. The interactions between individuals are not random, they follow predefined or evolved patterns. Then the {\em communication} between them is established. For a system that supports evolution, individuals have the ability to change, which enables the fittest is reserved, and the {\em adaptability} is possible.

A system is a set of interdependent and interacting modules (components or individuals), which forms an integrated whole. If partitioned modules are unrelated, the system is static and inactive, shows no interaction. In a dynamic, autonomous or adaptive system, the interconnectivity between modules is required. The separableness ensures that a system can be partitioned into modules, so each module can be designed or developed independently. It is the premise for concurrency, parallelism and distribution. Modules of a system can be connected by structural, functional, behavioural relationships. The connection bands partitioned modules together. Then the interaction between modules would occur frequently. Many interactions are not random. They will show some kind of patterns (e.g., protocols and laws in the Internet and the biosphere). These patterns are the manufactured products of the communication, which makes them coordinate, compete, develop and struggle for existence. The connection supports communications. In traditional paradigms of the digital world, the communication is supported by limited approaches, e.g., semaphore, shared memory. They are mainly designed for exchanging data between partitioned modules in a system. No valid channel is provided for every object that they can communicate without any constraints.

Motivated by issues discussed above, a protocol programming paradigm is presented for designing and implementing a system. Protocols are predefined communicating patterns. They are used to connect and coordinate each module in a system. In this paradigm, programming is divided into two aspects: protocol implementation and domain implementation. Protocol implementation provides a platform for connecting or coordinating instantiated objects. Domain implementation can use the traditional methods to encapsulate objects. But supported by a protocol implementation, when designing and encoding a system, the development process become simple. Because the communication can be designed to be connectionless, modules can have the ability to update itself without stopping the system.

In order to show the methodology of protocol programming, next, we give a simple discussion about the implementation developed in our work. 

\section*{Protocol Implementation}
\label{sec:protocol-implementation}

\begin{figure}[h]
	\centering
	\includegraphics[width=9cm]{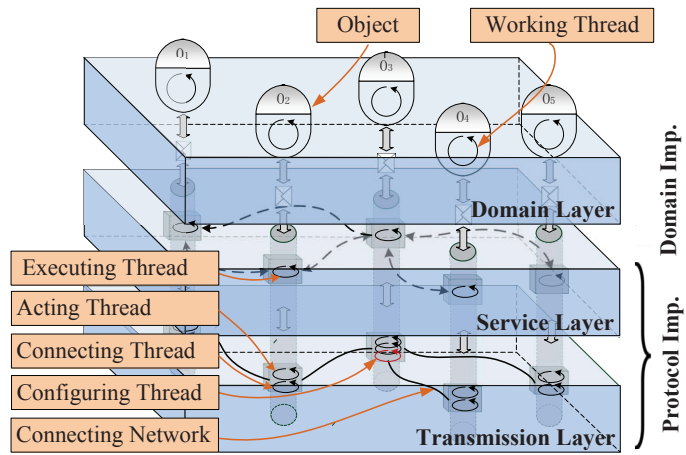}
	\caption{Protocol Programming Architecture}
	\label{fig:protocol_programming_paradigm}
\end{figure}

Our protocol implementation has two layers: {\em transmission layer} and {\em service layer}. The transmission layer supports the connection, providing a channel for communication. In this layer, a protocol is designed to establish a thread network, referred to as the connecting network. The service layer supports the connected objects to register, require and execute a service according to a defined protocol. The framework about the programming paradigm is shown in Figure \ref{fig:protocol_programming_paradigm}.

The protocol implementation is divided into different layers. Each layer is implemented by a class. If a class implements a non-bottom layer, this class should inherit from a parent class that implements a lower layer. It is better that each layer is designed to be transparent, only the interface is known by the upper layer the same as the OSI model \cite{zimmermann1980osi}. Every object instantiated from the transmission layer is connected into the connecting network, where all messages between objects are transmitted automatically. In the service layer, the service configuration, registration, request and execution are performed. In the domain layer, each object implements specific work. Next, the paradigm is discussed.

In the protocol implementation, the connection and communication are done by transmitting messages from one thread to another. Unlike transmitting an IP package through the Internet, in a local environment, all messages share the same memory space. Therefore, only the address of message is posted between threads. The structure of a message is given as follows, where $\text{\bf PMessage}$ is a class. Every message is an instantiation of this class or its sub-class.
\[
\begin{array}{l}
  \text{PMessage}\{  \\
  \qquad \text{LAYER}  \quad m\_Layer;	  \\
  \qquad \text{ACTION}  \quad m\_Action;	  \\
  \qquad \text{ADD}  \quad m\_Intend;	  \\
  \qquad \text{ADD} \quad  m\_Creator;	  \\
  \qquad \text{bool} \quad  m\_NeedEchoFlag;\\
  \qquad \text{size\_t} \quad  m\_TimeToLive;	  \\
  \qquad \text{TIME} \quad  m\_CreateTime;	  \\
  \qquad \text{vector} \langle \text{ADD} \rangle \quad  m\_Route\_v;	  \\
  \qquad \text{vector} \langle \text{string} \rangle \quad  m\_Param\_v;\}	  \\
\end{array}
\]
The $\text{LAYER}$ and $\text{ACTION}$ are enumerated data types. The field $m\_Layer$ is used to indicate which layer a message should be delivered for further processing. Three types are defined, representing each layer respectively. Each layer supports a certain numbers of actions suggested in the field $m\_Action$. The $ADD$ represents the address (or identification) of an object. The fields $m\_Intend$ and $m\_Creator$ are the destination and creator of a message. In the field $m\_NeedEchoFlag$, a $true$ value indicates that a response is required by $m\_Creator$. The $m\_TimeToLive$ is a hop count used to limit a message's lifetime. The $m\_CreateTime$ can be used to line up arrived messages for avoiding problems such as data access violation when accessing a critical resource. When a message is transmitted through the network, the field $m\_Route\_v$ can be used to record the route information. The field $m\_Param\_v$ stores parameters of message.

\subsection*{Transmission Layer}

The main purpose of the transmission layer is to build an infrastructure for message transmission. We design a class named as $\text{\bf PTrans}$ to do this. Three threads are created in each instantiated object of $\text{\bf PTrans}$: configuring thread, connecting thread and acting thread, used for configuration, connection and action respectively. They run concurrently.Five kinds of actions are defined in the transmission layer: $\text{CONFIG}$, $\text{HELLO}$, $\text{ECHO}$, $\text{TICK}$ and $\text{EXIT}$.

All messages are posted through the connecting thread. A connecting thread retrieves messages from its message queue. If a message's destination matches itself, the message is taken off. Otherwise, it is retransmitted according to a routing protocol. Because other messages in its message queue may be waiting for transmitting, it is not reasonable to spend too much time for processing a message in the connecting thread. The connecting thread pushes its message into a transmitting message list and return immediately. All messages in the list are processed by the acting thread. If the message list is empty, the acting thread is blocked. If a message is added, the connecting thread sets a semaphore. Then the acting thread is waken up to process the arrived message.

In protocol programming, the most important problem is how to make every object connected. The topology of connecting network determines the strategy to establish it. In our implementation, we design a tree structure. The object in the structure also is referred to as {\em node}. The $Master$ is the {\em root} of the structure and a {\em leaf node} is an object without any child. In our implementation, every object is identified by the connecting thread handle. The child object (the child) is used as a member variable of a parent object (the parent). The child has its parent's connecting handle as a parameter for initialization. Then, after all objects are initialized, except the root object, every object knows its parent's connecting handle.

After a child is initialized, using parent's connecting handle, the child reports its connecting handle to its parent by posting a $\text{CONFIG}$ message, then waits for a response. The parent collects the child's identification from the message and posts a message in return. After this process, each pair of parent and child know the connecting threads between each other. All configuring threads exist except that in the $Master$, which will be running until the system terminates.

The configuring thread in $Master$ broadcasts a $\text{HELLO}$ message to all its children to start a connecting process. The $Master$ will wait until all responses of its children are returned. After a $\text{HELLO}$ is received from its parent, every object also broadcasts a $\text{HELLO}$ to its children. It responds an $\text{ECHO}$ to its parent in two conditions: all feedbacks from children are returned or it has not child at all. In a concurrent environment, some threads may be blocked or suspended, if a time-out error occurs,, the $Master$ can restart a new connecting process. When children's $\text{ECHO}$ arrive at the $Master$, it means that every object is instantiated successfully. A connecting tree structure is established.

All messages are transmitted in the connecting network. If there is no message to process, a connecting thread is blocked by calling a {\em GetMessage(\&msg, ...)} function. A blocked connecting thread is aroused by arrived messages. Because all threads are running concurrently that some threads may be jammed or blocked, the $\text{TICK}$ message is used to check the state of an object. The $\text{EXIT}$ is used for system to exit. Every node retransmits this message to its children. All nodes from the leaves to the root will release its resources and send a response to its parent. 

It the arrived message does not belong to the transmission layer, the acting thread calls a pure virtual function {\em Trans\_Outlet} to process it.

\subsubsection*{Interface}

The connecting thread and the {\em Trans\_Outlet} are the interface between the transmission layer and the service layer. In the service layer, a requirement is submitted by using the handle as parameter to post messages, and the {\em Trans\_Outlet} is overridden to provide services.

\subsection*{Service Layer}
In this paper, a domain service (or a service for short) is defined as a function that can be executed to support a request in an application. It is provided in the domain layer. The service layer doesn't provide any domain service. The major function of the service layer is to supervise the services provided in the domain layer.

The service layer is implemented by a $\text{\bf PService}$ class inherited from the $\text{\bf PTrans}$. The pure virtual function {\em Trans\_Outlet} in transmission layer is overridden in the $\text{\bf PService}$ class. Because the $\text{\bf PTrans}$ is inherited, any object instantiated from the $\text{\bf PService}$ class is automatically connected into the connecting network. If a matched message cannot be processed in the transmission layer, the function {\em Trans\_Outlet} will be called. Three actions are defined in the service layer: $\text{REG}$, $\text{REQ}$ and $\text{ACK}$. 

Because the implementation of service layer is transparent and independent from the domain service. Therefore, the service layer should be given the information about a domain service. A $\text{\bf PServicedepict}$ class is defined to contain this information, named as the service description. Each service must have a service description. The content of the service description is introduced next.

The service register is started by the $Master$ through broadcasting a $\text{REG}$ message to its children. The $\text{REG}$ message is spread from the root node to all leaf nodes. Every node collects service descriptions from its children and reports all services it knows to its parent. When the service description transmits through the connecting network, routing information is collected. With this protocol, every object contains service descriptions about it and its children. The $Master$ knows the all services in a system.

If a service is requested by an object, it is done by packaging a $\text{REQ}$ message into the connecting network, where the name of the service and possible parameters are filled. In our implementation, a string matching strategy is used to match a service. When requesting a service, the synchronization and asynchronization should be considered. In asynchronously calling, the object keeps going after a requirement was posted. Otherwise, it is blocked to wait an $\text{ACK}$ message. The message is packaged by calling a requesting function defined in the $\text{\bf PService}$ class, then transmitted from the requesting object towards the $Master$. If an object has a service description matching this requirement, the message is retransmitted to the provider directly by routing information in the service description.

If a $\text{REQ}$ message arrives at its provider, the message is appended into a local message list of the $\text{\bf PService}$ class, named as the service message list. An executing thread defined in the class is always waiting for processing each message respectively. Two approaches can be used to implement a service: a thread or a function. Their addresses are stored in the service description. In the service description, two fields are declared to indicate which approach is supported and whether or not the service is reenterable.

Many mechanisms can be designed to support the execution of a service. For example, for a specific service, if the service is executed by a thread, then the service message can be posted into the thread directly. If a service is declared in a class in the domain layer, then for each arrived message, a new object of the class can be instantiated to handle this service. If the codes are reenterable, it is possible to create a new thread to process each message separately and concurrently. Furthermore, the dynamic scheduling can be designed to support the load balancing. These mechanisms are automatically supervised in the service layer. Only the service descriptions are shared between the service layer and the domain layer.

After a $\text{REQ}$ message is executed, if a response is required in the synchronously calling, it will return an $\text{ACK}$ message.

\subsubsection*{Interface}
One interface of the service layer is the requesting function, where the $\text{REQ}$ message is generated and posted into the connecting network. Another interface is a service description list. It is a member variable of the $\text{\bf PService}$ class used by the domain layer to report its services.

\section*{Domain Implementation}
\label{sec:domain-implementation}

In domain implementation, the traditional programming methods can be used to provide the domain services. However, by inheriting a protocol implementation, all instantiated objects are connected into the connecting network. Every connected object can communicate with each other by predefined protocols. Therefore, the difficulty to partition a problem into discrete tasks is reduced. Following the predefined protocol, every  module can be developed independently. Because the concurrency, distribution, or load balancing, etc. can be supported transparently by the protocol implementation, it also reduces the burden to develop the domain implementation.

\section*{Evaluation}

Three classes, inherited from the $\text{\bf PService}$, are declared in the domain layer. Each provides a toy service, named as $Service1$, $Service2$ and $Service3$. They separately execute $1\times 10^6$, $2\times 10^6$ and $1\times 10^6$ Floating Multiply Instructions (FMI). Three objects are instantiated from each class respectively. They are connected into a connecting network with a $Master$ object instantiated from the $\text{\bf PService}$ class.

\begin{figure}[h]%htbp[width=9cm]
\centering
\subfigure[Time Consuming]{
\begin{minipage}[h]{0.48\textwidth}
\includegraphics[width=7cm]{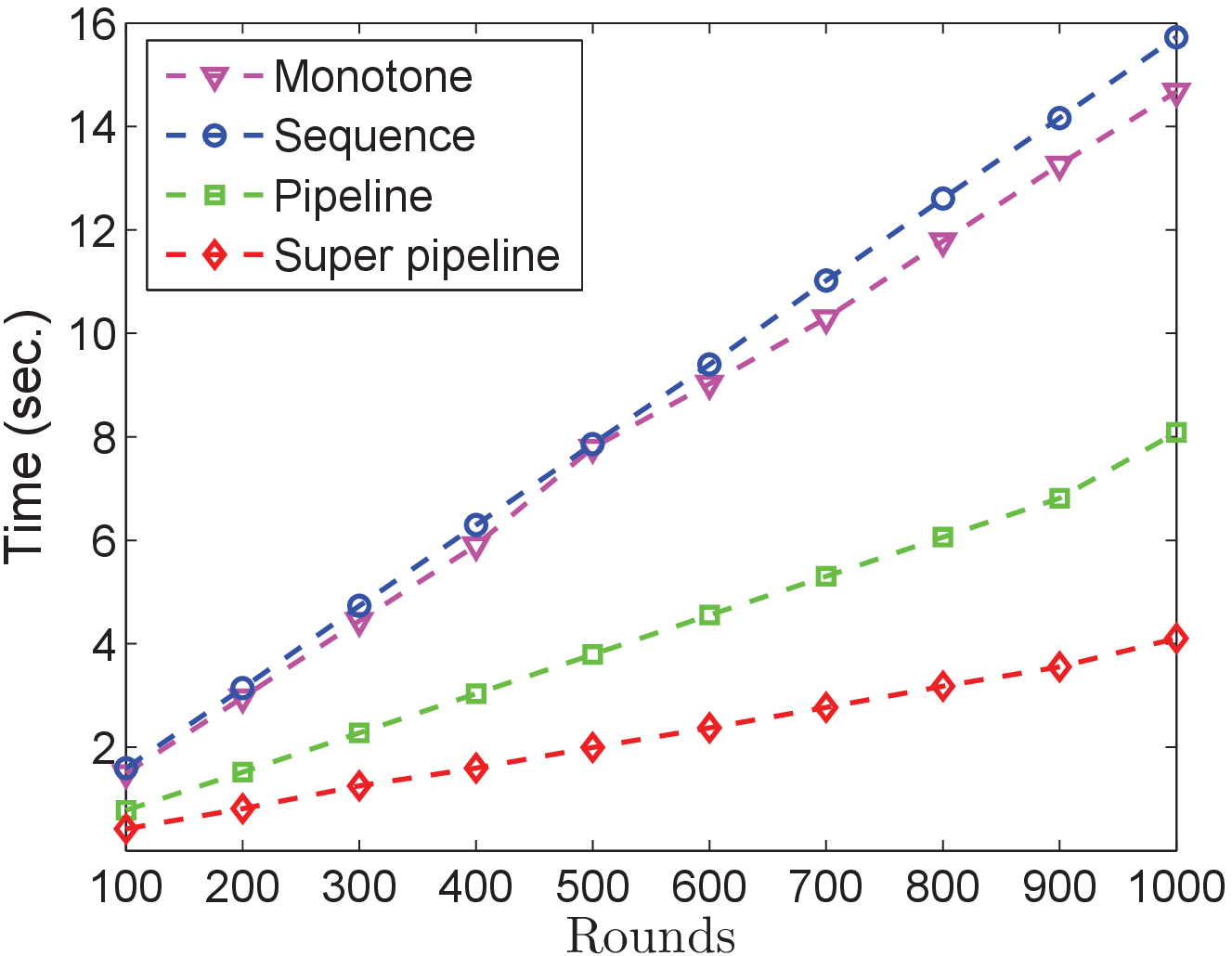}
\centering
\label{subfig:a}
\end{minipage}
}
\subfigure[Speedup of Super Pipeline]{
\begin{minipage}[h]{0.48\textwidth}
\includegraphics[width=7cm]{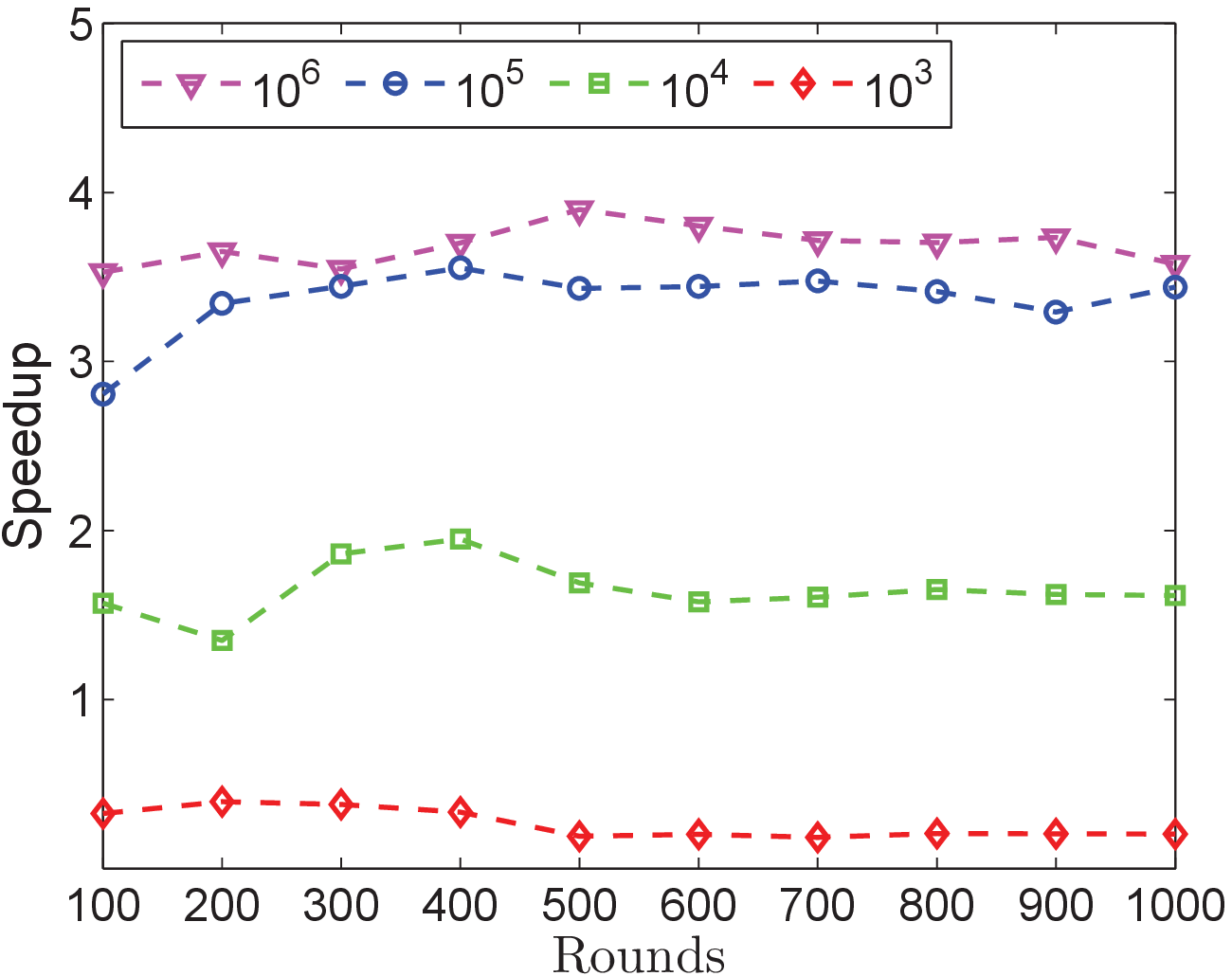}
\centering
\label{subfig:b}
\end{minipage}
}
\caption{Evaluation of Protocol Programming}
\label{fig:evaluation}
\end{figure}

In the first experiment, the monotone model runs every service in a single function without transmitting message. In the sequence model, the $Master$ orderly executes each service by synchronously posting the $\text{REQ}$ message to the three objects. In the pipeline model, the objects are organized in a line and asynchronously called from one to another. They are executed simultaneously. Because the computation complexity of the $Service2$ is doubled, it becomes the bottleneck in the pipeline model. In the super pipeline, two objects are instantiated to provide the $Service2$ service. Running time\footnote{The evaluation uses a Window 7 OS with Intel i5-2400 3.10GHz 4 Core CPU and a 8 GB memory.} of the four models is shown in Figure \ref{subfig:a}.

In Figure \ref{subfig:a}, the performance of the monotone model can be seen as a benchmark. In the sequence model, because synchronously calling is used, the time consuming is increased due to message transmitting. In the pipeline model, the expense caused by transmitting messages can be offset by allowing the services to be processed simultaneously. The super pipeline has a load balancing mechanism, therefore, the highest performance is received.

In Figure \ref{subfig:b}, the speedup\footnote{It is computed by dividing the time consuming in the monotone model and the super pipeline model.} of the super pipeline model is compared using different numbers of FMI. When the number of FMI is decreased exponentially in each service, the speedup is decreased accordingly. When the number of FMI is decreased to $10^3$, the cost of transmitting message negated the superiority of the super pipeline. Therefore, the speedup is less than 1.

The implementation has a very high speedup, because, instead of a single thread, multi threads are used to implement the services. The main advantage of protocol programming is that the concurrency is managed automatically by the protocol implementation.

\section*{Discussion}
\label{sec:discussion}

In protocol programming, three types of objects exist: the $Master$ object, the $Server$ object and the $Router$ object. The $Master$ is designed for governing a system. It is also helpful to use multi-master objects in a system. Each performs a specific assignment. The $Server$ refers to the object who provides the service. A $Router$ is a node, which only transmits messages and no service or management is assigned. It can can be used to organize a system or partition the message space. In a system, mixed objects can be used.

After a system is initialized, every object inherited from the $\text{\bf PService}$ class will have at least three threads. All communications between them are performed by message transmitting. Therefore, the concurrency is naturally supported. When a system is running, objects can be added or erased from the connecting network dynamically, which can be used to support the dynamic scheduling and the load balancing. 

Because the connecting network provides a connectionless service to all objects, the service provided by the program can be updated without stopping the system. Other methods developed in the Internet protocol can be introduced for the protocol implementation, e.g., domain name service, error control, structured naming strategy, etc. In addition to keeping the program and the service description locally, Both can be registered and stored in a remote server, then accessed though the Internet. If a service is supported in a remote system, the message can be transmitted across the Internet and executed in the remote system. If possible, the program can also be downloaded directly and run locally.

In summary, the advantages of the protocol programming include: (a) a novel paradigm for programming is presented, which divides the process into two aspects: protocol implementation and domain implementation. (b) the connection as a mechanism to support communication for programming is emphasized and developed. (c) protocols are designed to control the communication between objects, which enables the communications across heterogeneous programs, systems and the Internet. (d) Because one protocol implementation can be shared by different applications, it ensures a broad range of code reuse. (e) mechanism such as concurrency, parallelism, distribution, pipeline and adaptability, etc. can be governed transparently, autonomously, even adaptively.  (g) it supports the autonomy of objects, which enables an object to update itself when the system is running.

\bibliography{reference}

\bibliographystyle{Science}

\end{document}